\documentclass[mathleft,fleqn,%
]{an}
\usepackage{amsmath}
\usepackage{graphicx}
\usepackage[varg]{txfonts}
\usepackage{natbib}
\bibpunct{(}{)}{;}{a}{}{,}
\begin{document}

\Pagespan{1}{}
\Yearpublication{2014}%
\Yearsubmission{2014}%
\Month{0}%
\Volume{999}%
\Issue{0}%
\DOI{asna.201400000}%

\received{2017 Feb 10}
\accepted{2017 Jul 29}
\publonline{XXXX}

\title{Analytic construction of baroclinic tori by non-linear summation}
\titlerunning{Analytic construction of baroclinic tori by non-linear summation}
\author{D.\,N. Razdoburdin\inst{1,2}\fnmsep\thanks{Corresponding author:
        {d.razdoburdin@gmail.com}}}

\institute{
Sternberg Astronomical Institute, Moscow M.V. Lomonosov State University, Universitetskij pr., 13, Moscow 119992, Russia
\and 
Department of Physics, Moscow M.V. Lomonosov State University, Moscow, 119992, Russia
}


\defcitealias{amendt-1989}{ALA89}

\abstract
{
In this paper analytical method for constructing baroclinic flow is presented.
Presented method allows the construction of baroclinic tori with  significant vertical shear, which are stable by the Solberg--H{\o}iland criteria.
The presented method can be useful in the investigations of the influence of background structure on vertical-shear- and entropy-driven instabilities.
}
\maketitle

\section{Introduction}
Angular momentum transfer to the periphery of astrophysical disks is one of the long-standing problems.
However it remains relevant even today.

Turbulence driven by magneto-rotational instability (MRI) (based on the Velikhov–Chandrasekhar instability, \cite{velikhov-1959}; \cite{chandrasekhar-1960}) discovered by \cite{balbus-hawley-1991}, does not occur in weakly ionized parts of disks.
An example of such parts is the dead zones in protoplanetary disks (see \cite{armitage-2011} and \cite{turner-2014} for detailed reviews).

Moreover, the $\alpha$-parameter of \cite{shakura-sunyaev-1973} provided by direct numerical simulation of MRI-driven turbulence does not exceed a value $\alpha\sim 0.03$ (\cite{simon-2012}).
At the same time, the interpretation of accretion disks observations indicates that $\alpha>0.1$ (\cite{king-2007}, \cite{suleimanov-2008}, \cite{kotko-lasota-2012}, \cite{lipunova-2015}, \cite{lipunova-malanchev-2016}).

Therefore alternative mechanisms of angular momentum transfer have to be sought.

One of such alternatives is a subcritical transition to turbulence through the transient growth of linear perturbations (see the review by \cite{razdoburdin-zhuravlev-2015} and references therein for more details).
Unfortunately, numerical simulations in local approach do not show any subcritical transition in homogeneous, hydrodynamic, Keplerian flows (see \cite{shen-2006}).
However, this negative result may have been due to insufficient numerical resolution (see discussion in \cite{mukhopadhyay-2005} and \cite{razdoburdin-zhuravlev-2017}).

Another alternative scenario is based on the instabilities that occur in stratified flows.
This sort of instability is driven by the entropy gradient.
First evidence for its existence was found by \cite{klahr-bodenheimer-2003} in global simulations of disk with radial entropy gradient.
The authors called it a ''global baroclinic instability'' (GBI).
However, the subsequent studies made by \cite{klahr-2004} and \cite{johnson-gammie-2005a} didn not find any linear instability in the framework of the inviscid dynamics.
Further, \cite{johnson-gammie-2006} didn not find any instability in the shearing box model employing a nonlinear approach.

New impetus to entropy-driven instability was given in papers by \cite{petersen-2007a} and \cite{petersen-2007b}.
In contrast to \cite{johnson-gammie-2006}, the authors took cooling function into account and used the spectral approach.
They found a spontaneous formation of long-lived vortices, which produce density waves.
Finally, \cite{lesur-papaloizou-2010} observed this instability in the local approach without any vertical stratification.
The rate of angular momentum transfer by density-waves corresponds to an the effective $\alpha \sim 10^{-3}$.
The instability was found to be subcritical.
This means that the instability operates only for finite amplitude perturbations with some threshold of their initial amplitude.
Because of its local and subcritical nature, this instability is called now a ''subcritical baroclinic instability'' (SBI).
The generation of perturbations with finite amplitude that gives rise to SBI was investigated in \cite{klahr-hubbard-2014} and \cite{lyra-2014}.

In presence of critical layers in stratified flow, zombie vortex instability (ZVI) can occure.
Critical layers roll up into vortices, which give birth to other critical layers, which in turn produce new vortices.
The corresponding angular momentum transfer was found to be $\alpha \sim 10^{-3}$
(see \cite{marcus-2013}, \cite{marcus-2015}, \cite{marcus-2016} for details).

Another kind of instability occurs in baroclinic flows.
Baroclinic flow is the most general case of the rotating flow. 
In contrast with more particular case of barotropic flow, in the baroclinic one the pressure cannot be represented as function of density alone.
This means that cross product $\nabla p \times \nabla \rho \ne 0$.
According to the Poincare-Wavre theorem, this means that the angular velocity of the fluid, $\Omega$, is the function of both radial and vertical coordinates (see \cite{tassoul-1978} for details).

Baroclinic flows are unstable owing to the so-called vertical shear or Goldreich -- Schubert -- Fricke (GSF) instability
\footnote{in some papers it is also called the vertical shear instability (VSI).}.
It was first discovered independently in the papers by \cite{goldreich-schubert-1967} and \cite{fricke-1968}.
This instability is driven by the combination of vertical gradient of angular velocity (vertical shear) and heat transport.
Linear and nonlinear regimes of this instability were investigated in \cite{urpin-brandenburg-1998}, \cite{urpin-2003}, \cite{arlt-urpin-2004} in the context of disk physics.
But only in \cite{nelson-2013} a clear evidence for turbulence and the associated angular momentum transport was found.
It was shown that GSF instability can provide a transition to turbulence in locally isothermal regime with the effective viscosity $\alpha\sim 10^{-3}$ similar to the value provided by SBI and ZVI (the most recent simulation has been performed by \cite{richard-2016}).

One of the important issues is the influence of the background structure on the properties of GSF-instability.
In \cite{barker-latter-2015}, the linear regime of this instability was investigated for both ''locally isothermal'' (as in \cite{nelson-2013}) and ''locally polytropic'' background models.
The challenge of such investigations is in the construction of stationary baroclinic background.
Since baroclinic flows are the most general case of stratified flow, their apprehensible construction is of interest in the context of the investigation how the background 
affects the SBI and ZVI, as well as the GSF instabilities.

Baroclinic flows cannot be constructed by analytical integration of equilibrium equations with a given rotation law.
That is why all previous studies (starting from \cite{frank-robertson-1988}) have constructed the baroclinic background by setting temperature, entropy, pressure or density as functions of coordinates and the subsequent determination of angular velocity from the equations of the dynamic equilibrium.
The shortcoming of such a method is the need to postulate a vague two-dimensional function which fully determines the properties of the flow.
And dependence of properties of the flow on this function is non obvious.
Therefore, specification of such a preset function, which would correspond to the flow with the required properties, is a nontrivial task.

In this paper an alternative method for the construction of the analytical stationary baroclinic flow is presented (however, it can be reduced to one described above).

This method allows one to construct baroclinic tori with a significant rate of baroclinicity as well as zero density and pressure at the boundary.
The latter can be important for the studies by vertically global shearing box (see \cite{mcnally-pessah-2015} about this approach).
The baroclinicity of the flow can be easily changed by varying the inner parameters of construction.

This method is based on the nonlinear summation of barotropic solutions which has been first investigated in \cite{amendt-1989} (hereafter \citetalias{amendt-1989}).
However, the original method does not provide the coincidence of zero-pressure and zero-density surfaces.
More precisely, the zero-pressure surface lies within the zero-density surface.
In combination with the equation of state for an ideal gas, it results in the entropy equal to minus infinity at the boundary.
Since in the model by \citetalias{amendt-1989} both pressure and entropy grow out of the boundary of the torus, this results in the destabilization of the flow with respect to axisymmetric adiabatic perturbations.
That is why the original method does not allow the construction of Solberg--H{\o}iland stable flows.

We will show in the following that the modification of \citetalias{amendt-1989} method allows us to construct baroclinic flow with coinciding surfaces of zero pressure and zero density. 
In such a flow entropy is finite and it is stable by Solberg--H{\o}iland criterion.

\section{Construction of baroclinic flow}

Stationary axisymmetric flow in the Newtonian potential of the central body obeys the following set of equations:

\begin{equation}
\label{initial_r}
\frac{\partial p}{\partial r}=-\frac{\rho r GM}{\left(r^2+z^2\right)^{3/2}}+\rho r \Omega^2
\end{equation}

\begin{equation}
\label{initial_z}
\frac{\partial p}{\partial z}=-\frac{\rho z GM}{\left(r^2+z^2\right)^{3/2}}
\end{equation}
Here, $M$ is mass of the central body, $G$ is gravitational constant and $r$ and $z$ are the cylindrical coordinates.

For parametrization of the problem, the parameter $r_0$ is chosen.
It is equal to the distance at which the pressure reaches its maximum value on the equatorial plane.
The rotation angular velocity at $r=r_0$ is denoted as $\Omega_0=\Omega(r=r_0, z=0)$.
Thus, $\partial p/ \partial r=0$ at $r=r_0$, $z=0$, whereas $GM=r_0^3 \Omega_0^2$.

So, the set of hydrodynamic equations take the form:
\begin{equation}
\label{Euler_r}
\frac{\partial p}{\partial r}=-\frac{\rho r \Omega_0^2 r_0^3}{\left(r^2+z^2\right)^{3/2}}+\rho r \Omega^2
\end{equation}

\begin{equation}
\label{Euler_z}
\frac{\partial p}{\partial z}=-\frac{\rho z \Omega_0^2 r_0^3}{\left(r^2+z^2\right)^{3/2}}
\end{equation}
Equations  (\ref{Euler_r}) and (\ref{Euler_z}) contain three unknown functions: $\rho$, $p$ and $\Omega$, and can not be solved without any additional assumptions.

\subsection{Barotropic solutions}
First, let us solve this set (following \cite{goldreich-1986}, \cite{kojima-1989}, \cite{kojima-1989b}, \cite{razdoburdin-zhuravlev-2012} and others) for the particular case of barotropic flow 
with $\Omega=\Omega_0\left(r/r_0\right)^{-q}$ and adiabatic equation of state $p=\rho^{\gamma}$.

With this simplifications, it is not difficult to obtain the solution of (\ref{Euler_r}) and (\ref{Euler_z}):
\begin{equation}
\label{rho_barotrop}
\rho=\left(\frac{\Omega_0^2r_0^2(\gamma-1)}{\gamma}\right)^{1/(\gamma-1)}\left(\frac{1}{\sqrt{r^2+z^2}}+\frac{r^{2-2q}}{2-2q}+C_q\right)^{1/(\gamma-1)}
\end{equation}

\begin{equation}
\label{p_barotrop}
\rho=\left(\frac{\Omega_0^2r_0^2(\gamma-1)}{\gamma}\right)^{\gamma/(\gamma-1)}\left(\frac{1}{\sqrt{r^2+z^2}}+\frac{r^{2-2q}}{2-2q}+C_q\right)^{\gamma/(\gamma-1)}
\end{equation}
Here, $C_q$ is an integration constant which can be easily found from the boundary condition $\rho(r=\{r_1,r_2\},z=0)=0$, where $r_1$ and $r_2$ are inner and outer radii of the flow.
\begin{equation}
\frac{1}{r_1}+\frac{r_1^{2-2q}}{2-2q}+C_q=0
\end{equation}

\begin{equation}
\frac{1}{r_2}+\frac{r_2^{2-2q}}{2-2q}+C_q=0
\end{equation}

This set contains the three unknowns: $r_1$, $r_2$ and $C_q$. 
Thus, the torus is fully determined by setting one of them.
However, it is more convenient to use the radial size of the torus $R_d=r_2-r_1$ as the parameter.
So, the final set for $C_q$, $r_1$ and $r_2$ take the following form:
\begin{equation}
(2-2q)+r_1^{3-2q}+(2-2q)C_qr_1=0
\end{equation}

\begin{equation}
(2-2q)+r_2^{3-2q}+(2-2q)C_qr_2=0
\end{equation}

\begin{equation}
r_2=R_d+r_1
\end{equation}
Thus, the barotropic solution of the set (\ref{Euler_r}) and (\ref{Euler_z}) is characterised by the following parameters: $\gamma$, $R_d$, $r_0$, $q$ and $\Omega_0$.
In this paper the parameters $\gamma$ and $\Omega_0$ are specified by  $\gamma=5/3$ and $\Omega_0=1$.

\subsection{Construction of barocline}
\label{construction}
As was mentioned previously, the method for the construction of baroclinic flows is based on \citetalias{amendt-1989}.
However, the details of the construction differ significantly.

The method proposed by \citetalias{amendt-1989} has significant drawbacks.
One of them is a mismatch of the zero-density and the zero-pressure surfaces.
In combination with the equation of state for an ideal gas, it results in infinite entropy at the surface of the zero pressure (as long as density does not vanish there).
Another issue is the instability of the resulting barocline by the Solberg--H{\o}iland criterion.
Additionally, the method proposed by \citetalias{amendt-1989} does now allow the construction of highy baroclinic flows.
Note that here we suggest the corrections that fix all the above drawbacks.

Let $(\rho_1, p_1, \Omega_1)$ and $(\rho_2, p_2, \Omega_2)$ be two explicit barotropic solutions of the set (\ref{Euler_r}) and (\ref{Euler_z}).
Each solution is characterised by the same frequency $\Omega_0$ and adiabatic index $\gamma$.
However, their rotation index $q$, scale parameter $r_0$ and radial size $R_d$ differ.

Let us write the set of Euler equations for both of these solutions:
\begin{equation}
\label{setr1}
\frac{\partial p_1}{\partial r}=-\frac{\Omega_0^2 r_{01}^3 \rho_1 r}{(r^2+z^2)^{3/2}}+\Omega_0^2 r \rho_1 \left(\frac{r}{r_{01}}\right)^{-2q_1}
\end{equation}
\begin{equation}
\label{setz1}
\frac{\partial p_1}{\partial z}=-\frac{\Omega_0^2 r_{01}^3 \rho_1 z}{(r^2+z^2)^{3/2}}
\end{equation}
\begin{equation}
\label{setr2}
\frac{\partial p_2}{\partial r}=-\frac{\Omega_0^2 r_{02}^3 \rho_2 r}{(r^2+z^2)^{3/2}}+\Omega_0^2 r \rho_2 \left(\frac{r}{r_{02}}\right)^{-2q_2}
\end{equation}
\begin{equation}
\label{setz2}
\frac{\partial p_2}{\partial z}=-\frac{\Omega_0^2 r_{02}^3 \rho_2 z}{(r^2+z^2)^{3/2}}
\end{equation}

Now let's multiply (\ref{setr1}) and (\ref{setz1}) by some real number $\beta$, and (\ref{setr2}) and (\ref{setz2}) by the $1-\beta$.
Then, we add (\ref{setr1}) to (\ref{setr2}) and equation (\ref{setz1}) to (\ref{setz2}):

\begin{equation}
\begin{aligned}
\label{setbr}
&\frac{\partial(\beta p_1 + (1-\beta)p_2)}{\partial r}=-\frac{\Omega_0^2 r \left(\beta r_{01}^3\rho_1+(1-\beta)r_{02}^3\rho_2\right)}{(r^2+z^2)^{3/2}}+\\
&+\Omega_0^2r\left[\beta\rho_1\left(\frac{r}{r_{01}}\right)^{-2q_1}+(1-\beta)\rho_2\left(\frac{r}{r_{02}}\right)^{-2q_2}\right]
\end{aligned}
\end{equation}

\begin{equation}
\label{setbz}
\frac{\partial(\beta p_1 + (1-\beta)p_2)}{\partial z}=-\frac{\Omega_0^2 z \left(\beta r_{01}^3\rho_1+(1-\beta)r_{02}^3\rho_2\right)}{(r^2+z^2)^{3/2}}
\end{equation}

Now, if the new functions 
\begin{equation}
\label{pb}
p=\beta p_1 + (1-\beta) p_2
\end{equation}
\begin{equation}
\label{rhob}
\rho=\frac{\beta r_{01}^3\rho_1 + (1-\beta) r_{02}^3 \rho_2}{r_{01}^3}
\end{equation}
\begin{equation}
\label{Omegab}
\Omega^2=\Omega_0^2 \frac{\beta \rho_1 \left(\frac{r}{r_{01}}\right)^{-2q_1} + (1-\beta) \rho_2 \left(\frac{r}{r_{02}}\right)^{-2q_2}}{\rho}
\end{equation}
are defined, (\ref{setbr}) and (\ref{setbz}) take the form:

\begin{equation}
\label{barocline_r}
\frac{\partial p}{\partial r}=-\frac{\rho r \Omega_0^2 r_{01}^3}{(r^2+z^2)^{3/2}} + r \rho \Omega^2
\end{equation}
\begin{equation}
\label{barocline_z}
\frac{\partial p}{\partial z}=-\frac{\rho z \Omega_0^2 r_{01}^3}{(r^2+z^2)^{3/2}}
\end{equation}
Equations (\ref{barocline_r}) and (\ref{barocline_z}) are exactly the same as the set (\ref{Euler_r}) and (\ref{Euler_z}).
It means that functions (\ref{pb})--(\ref{Omegab}) are the solutions of Euler equations.
Wherein $\Omega$ is functions of both coordinates $(r,z)$.
Thus, the solution describes baroclinic flow.

The problem of such a summation is a mismatch of the regions where the barotropic solutions are well defined (some regions lie inside the boundary of the first barotropic solution but outside the boundary of the second one).
In \citetalias{amendt-1989} pressure and density of barotropic solutions were redefined outside of the flow boundary to be negative.
This is the reason for the mismatch between the zero-pressure and zero-density surfaces in the original method.

\begin{figure}
\includegraphics[width=\linewidth]{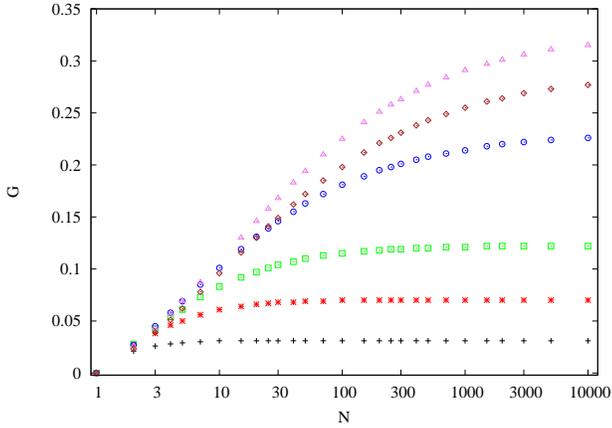}
\caption{
Dependence of global baroclinicity $G$ over the number of barotropes $N$ for different $k$ values.
Crosses correspond to $k=0$, double crosses to $k=-1$, squares to $k=-1.5$, circles to $k=-2$, triangles to $k=-2.5$ and diamonds to $k=-3$.
Sequences of $\{q_i\}$ and $\{\beta_i  \}$ are determined by (\ref{q_model}) and (\ref{beta_stable_model}).
$q_s=1.5$, $q_e=2$, $\gamma=5/3$, $R_d=1$.
}
\label{parametric}
\end{figure}

In this work the density and pressure of barotropic flows are defined to be equal to zero outside their boundaries.
Since the barotropic solution (\ref{rho_barotrop}) and (\ref{p_barotrop}) have spatial derivatives vanishing at the boundary, such an extension is smooth everywhere.

Unfortunately, the resulting baroclinic flow may acquire an infinite radial derivative of its thickness , $dH/dr=\infty$, in the regions, where the barotropic boundaries intersect with each other.
To solve this problems barotropic flows are chosen to lie one inside another like the matryoshka doll.
For simplicity, $r_{02}$ and $R_{d2}$ are chosen in such a way, that the inner and outer radii of both barotropes coincide.

The baroclinic solution is determined by $\beta$, $q_1$, $q_2$, $r_{01}$, $R_{d1}$, $\gamma$ and $\Omega_0$.
Hereafter, only the flows with $\Omega_0=1$, $\gamma=5/3$, $r_{01}=1$ and $R_{d1}=1$ will be considered.

Natural generalisation of the method is reduced to the  summation of more than two barotropic solutions.
In this case, the baroclinic solution is determined by the following functions:
\begin{equation}
\label{pfull}
p=\sum\limits_{i=1}^{N}\beta_i p_i
\end{equation}

\begin{equation}
\label{rhofull}
\rho=\sum\limits_{i=1}^{N}\beta_i r_{0i}^3 \rho_i
\end{equation}

\begin{equation}
\label{Omegafull}
\Omega^2=\frac{\Omega_0^2}{\rho}\sum\limits_{i=1}^{N}\beta_i \rho_i \left(\frac{r}{r_{0i}}\right)^{-2q_i}
\end{equation}
With additional condition $\sum \beta_i =1$.

Because the inner and the outer radii of barotropes are matched, the barotropes with lower $q$ lay inside the barotrope with the highest $q$, so the thickness of the barocline is equal to the thikness of barotropr with the highest $q$ for any radius.
Since the density and pressure in baratropic solutions as well as their spatial derivatives vanish at the boundary, the resulting baroclinic solution is smooth everywhere inside their boundaries.

Note that the method does not naturally provide the positiveness of $\Omega^2$, $p$ and $\rho$.
Thus, sign of $\Omega(r,z)^2$, $p(r,z)$ and $\rho(r,z)$ must be checked for any baroclinic solution.

Construction of self-gravitating flow with non-zero vertical shear rate can also be in interest.
For that case, solutions with zero vertical shear rate can be calculated, for example, with method described in \cite{hure-hersant-2017}.
Constraction of baroclinic solution is similar to described above.
The only difference is that value of $\Omega_{0i}^2 r_{0i}^3$ must be the same for different barotropic solutions.

In present paper, we do not take self-gravitation into account.
Investigation of baroclinic self-gravitating flows is an is an interesting subject to future studies.

\begin{figure}
\includegraphics[width=\linewidth]{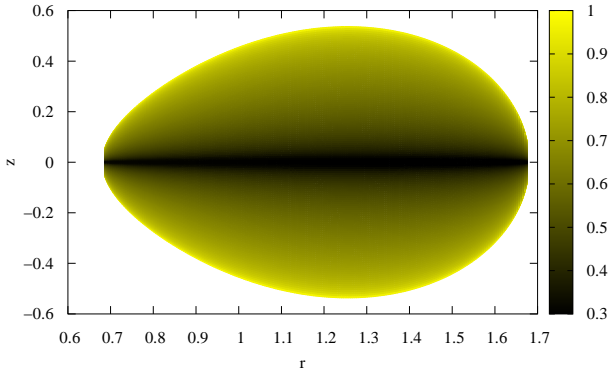}
\caption{
Entropy distribution in the $rz$-plane normalised by its maximum value in the flow.
The sequences of $\{q_i\}$ and $\{\beta_i  \}$ are determined by (\ref{q_model}) and (\ref{beta_stable_model}).
$q_s=1.5$, $q_e=2$, $N=249$, $k=-2.5$, $\gamma=5/3$, $R_d=1$.
The corresponding baroclinicity of the flow is $G=0.26$.
}
\label{entropy}
\end{figure}

\subsection{Solberg--H{\o}iland criterion}
The stability of baroclinic torus with respect to axisymmetric adiabatic perturbations is determined by the well-known Solberg--H{\o}iland criterion (see \cite{tassoul-1978}):
\begin{equation}
\label{HZ1}
\frac{1}{r^3}\frac{\partial j^2}{\partial r}-\frac{1}{\rho C_p}\nabla p \cdot \nabla S>0
\end{equation}

\begin{equation}
\label{HZ2}
-\frac{1}{\rho}\frac{\partial p}{\partial z}\left(\frac{\partial j^2}{\partial r}\frac{\partial S}{\partial z}-\frac{\partial j^2}{\partial z}\frac{\partial S}{\partial r}\right)>0
\end{equation}
It contains both the Rayleigh $d j^2/dr>0$ and the Schwarzschild $dS/dz>0$ conditions as particular cases.

Note that stability by Solberg--H{\o}iland criterion does not mean stability to all sorts of linear perturbations.
This criterion is valid only for axsysymmetric perturbations (see \cite{tassoul-1978}) and for perturbations with small radial and vertical scales combined with large azimuthal scale (see \cite{rudiger-2002}).
Full linear stability analysis can not be done with simple analytical criterion and require much more complicated investigations.
For example, in \cite{papaloizou-pringle-1984}, \cite{goldreich-1986}, \cite{kojima-1989}, such investigations were carried out for flows with constant entropy.
The role of entropy gradient was investigated, for example, in \cite{zhuravlev-shakura-2007b}.

\begin{figure}
\includegraphics[width=\linewidth]{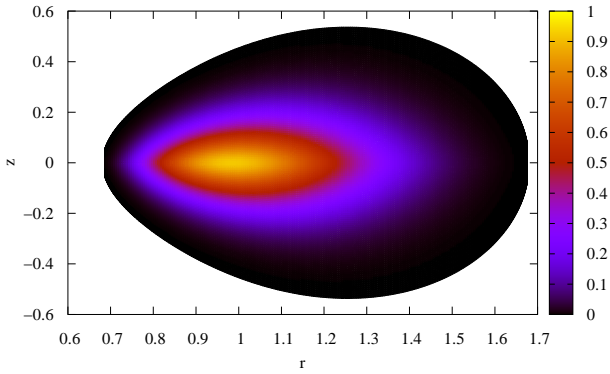}
\caption{
Pressure in the $rz$-plane normalised by its maximum value for the same parameters as in figure \ref{entropy}.
}
\label{p}
\end{figure}

\subsection{Measuring baroclinicity}
As in \citetalias{amendt-1989}, the following dimensionless parameter was used for quantitative characterisation of the baroclinicity:
\begin{equation}
\label{G}
G^2=\frac{1}{2}-\left(\frac{1}{2}\int \rho \frac{\nabla p \cdot \nabla \rho}{|\nabla p| |\nabla \rho|}drdz\right) \left(\int\rho dr dz\right)^{-1}
\end{equation}
In barotropic flows, pressure is a function of density only, that is why for this case the scalar product $(\nabla p \cdot \nabla \rho)=|\nabla p| |\nabla \rho|$ and $G=0$.
In contrast, to the extremely baroclinic case: $(\nabla p \cdot \nabla \rho)=-|\nabla p| |\nabla \rho|$ and $G=1$.

In order to see the regions which give the main contribution to full baroclinicity, the quantity $g$ can be introduced:
\begin{equation}
g^2=\frac{1}{2} - \frac{1}{2}\frac{\nabla p \cdot \nabla \rho}{|\nabla p| |\nabla \rho|}
\end{equation}
Obviously, $g(r,z)$ takes the values from zero in regions where baroclinicity is negligible up to unity in extremely baroclinic regions.

\subsection{Shear rates}
The common way to measure radial shear rate is to use the factor $q_r=-\frac{r}{\Omega} \frac{\partial\Omega}{\partial r}$.
It is equal to $q_r=3/2$ for Keplerian rotation and to $q_r=2$ for iso-momentum rotation.
In unstratified flows $q_r$ is a suitable indicator of stability since disks with $q_r>2$ are unstable according to the Rayleigh criterion.
However, in the case under consideration, the Rayleigh criterion is modified by the entropy gradient, see (\ref{HZ1}).
So, if entropy gradient has a stabilising effect the regions with $q_r>2$ can be stable.

The quantity $q_r$ represents the radial shear rate only.
In order to describe the vertical shear, let us first consider the shear rate tensor $\mathbf{\epsilon}_{ik}$,
Which is related to the stress tensor in the following way:
\begin{equation}
\mathbf{\sigma}_{ik}=\zeta (div~\mathbf{v}) \mathbf{I} + \eta \mathbf{\epsilon}_{ik},
\end{equation}
where $\mathbf{I}$ is a unit tensor, and $\eta$ and $\zeta$ are coefficients of dynamic and bulk viscosities.

For axisymmetric flow with the solely nonzero azimuthal velocity only, the two components of $\mathbf{\epsilon}_{jk}$ do not vanish:
\begin{equation}
\epsilon_{r\varphi}=r\frac{\partial \Omega}{\partial r}
\end{equation}
\begin{equation}
\epsilon_{z\varphi}=r\frac{\partial \Omega}{\partial z}
\end{equation}

Thus, $q_r$ is equal to the dimensionless $r\varphi$ component of the shear rate tensor $\epsilon_{ik}$ taken with a negative sign. 
Correspondingly, the natural form of vertical shear rate $q_z$ is the dimensionless $z\varphi$ component of $\epsilon_{ik}$ taken with the negative sign:
\begin{equation}
q_z=-\frac{\epsilon_{z\varphi}}{\Omega}=-\frac{r}{\Omega}\frac{\partial \Omega}{\partial z}.
\end{equation}

The distributions of $q_r(r,z)$ and $q_z(r,z)$ allow us to represent both radial and vertical shear rates in the flow.

\section{Results}
In order to show our method at work, several baroclinic flows are constructed.
For all of them the sequence $\{q_i\}$ is chosen in the form:
\begin{equation}
\label{q_model}
q_i=q_s+\frac{i \cdot dq}{N},
\end{equation}
where $i$ changes from $1$ up to $N-1$, $dq=(q_e-q_s)/(N+1)$.
This means that $q_i$ changes from $q_s+dq$ to $q_e-dq$ with step $dq$.

\begin{figure}
\includegraphics[width=\linewidth]{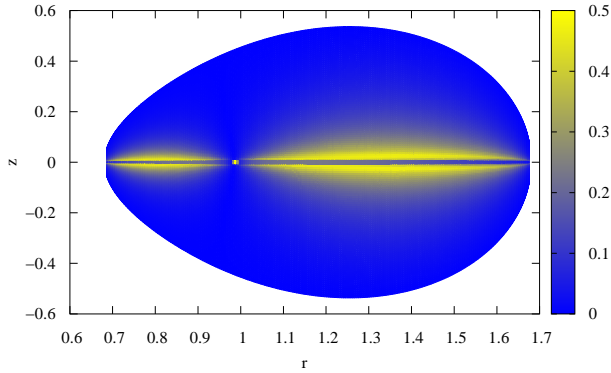}
\caption{
Local baroclinicity in the $rz$-plane for the same parameters as in figure \ref{entropy}.
}
\label{local_baroclinicity}
\end{figure}

First, let us construct  Solberg--H{\o}iland-stable baroclinic flow specifying the sequence of $\{\beta_i\}$ in the following way:
\begin{equation}
\label{beta_stable_model}
\beta_i=\frac{(i+1)^k}{\sum\limits_{i=1}^{N-1} (i+1)^k}
\end{equation}
The denominator in (\ref{beta_stable_model}) provides a unit sum over all coefficients $\sum \beta_i=1$.

The power $k$ specifies the relative weights of barotropic solutions.
For $k=0$, all barotropes have the same weight.
In figure (\ref{parametric}), we  show the dependence of global baroclinicity $G$ on the number of barotropes $N$ for different $k$.

It is found that $G$ increases monotonically up to some limit as $N$ goes to $\infty$.
In turn, this limit has maximum at $k=-2.5$.
Thus, the baroclinicity of the flow can be widely changed by changing $N$.
The highest baroclinicity that can be attained employing our method is $G=0.32$.
As can be seen, it exceeds the one obtained in \citetalias{amendt-1989} by more than an order of magnitude.

In order to demonstrate the profiles of physical values in the $rz$-plane, let us choose $q_s=1.5$, $q_e=2$, $N=249$, which correspond to $dq=0.002$.
Global baroclinicity of such a flow is equal to $G=0.26$.

Such value of $q_e$ results in significant difference of rotational profile from the Keplerian one.
However, for $q_e$ much closer to $1.5$ results do not significantly change.
For example, setting $q_e=1.55$ results in the slight decrease of $G$ only up to $G=0.18$.
Thus, our method allows to construct thin tori with significant baroclinicity too.

In figures (\ref{entropy}), (\ref{p}) and (\ref{local_baroclinicity}) the corresponding distributions of entropy, pressure and local baroclinicity $g$, respectively, are shown.
The pressure is maximum at the equatorial plane of the flow and decreases towards the boundary.
At the same time, the entropy is maximum right at the boundary.

\begin{figure}
\includegraphics[width=\linewidth]{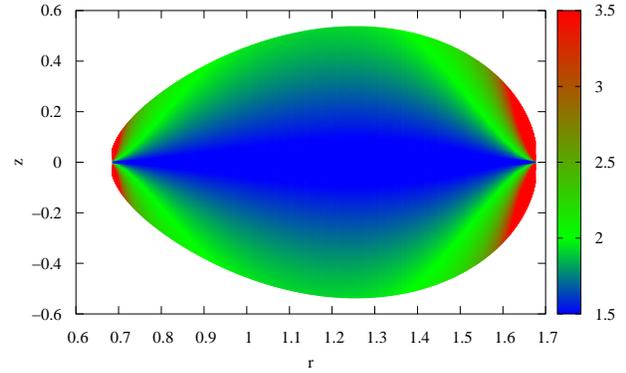}
\caption{
Radial shear rate $q_r$ in the $rz$-plane for the same parameters as in figure \ref{entropy}. Blue color corresponds to the regions with near-Keplerian rotation $q\sim 1.5$, green color corresponds to the regions with $q\sim 2$, and the red color corresponds to $q>2$. 
}
\label{local_q}
\end{figure}

\begin{figure}
\includegraphics[width=\linewidth]{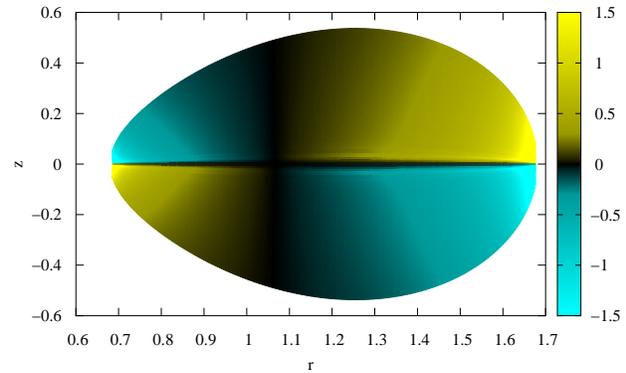}
\caption{
Vertical shear rate $q_z$ in the $rz$-plane for the same parameters as in figure \ref{entropy}.
}
\label{local_q_z}
\end{figure}

For unstratified flows with shear rate $q_r=-\frac{r}{\Omega} \frac{\partial\Omega}{\partial r}>2$ the specific angular momentum decreases with the radius; therefore  such flows are unstable according to the Rayleigh criterion.
However, in the presence of stratification the condition of the flow stability changes (see inequality (\ref{HZ1})).
Since entropy increases and pressure decreases to the periphery of the torus, the second term in (\ref{HZ1}) becomes negative.
This means that stratification stabilises the flow, and regions with $\partial j^2/\partial r<0$ become stable.
In figure \ref{local_q}, the distribution of radial shear rate in the $rz$-plane is shown.
In spite of the existence of areas with $q_r>2$, the flow is stable by the Solberg--H{\o}iland criteria.

Note that stability of regions with $q_r > 2$ does not require the flow to be baroclinic but only to be pseudo-barotropic (see \cite{tassoul-1978}).
That is, if entropy is a decreasing function of density, which is finite for zero density, regions with $q_r > 2$ will also stabilized by the entropy gradient.

In figure \ref{local_q_z}, the distribution of vertical shear rate in the $rz$-plane is shown.
The typical value of $q_z$ is some dozens.
So, the vertical shear is comparable to the radial shear.

The spatial distribution of $q_z$ has several features.
At first, it is antisymmetric with respect to the equatorial plane since $\Omega(z)$ is symmetric.
Second, $q_z$ vanishes at some radius inside the torus.
This result can be reproduced analytically by differentiating (\ref{Omegafull}) over $z$ taking into account what $r_{0i}$ of combined barotropes have the values close to each other.

In order to construct the Solberg--H{\o}iland-unstable barocline, the sequence $\{\beta_i\}$ can be chosen in the following form:
\begin{equation}
\label{beta_unstable_model_1}
\beta_i=\frac{(i+1)}{\sum\limits_{i=1}^{N-1} (i+1)-2.2\sum\limits_{i=141}^{199} (i+1)}
\end{equation}
for $1\le i \le 140$ and $200 \le i \le 249$, and
\begin{equation}
\label{beta_unstable_model_2}
\beta_i=-\frac{1.2\cdot(i+1)}{\sum\limits_{i=1}^{N-1} (i+1)-2.2\sum\limits_{i=141}^{199} (i+1)}
\end{equation}
for $140 < i < 200$.
As in (\ref{beta_stable_model}), here the denominator provides a unit sum of all coefficients $\sum \beta_i=1$.

\begin{figure}
\includegraphics[width=\linewidth]{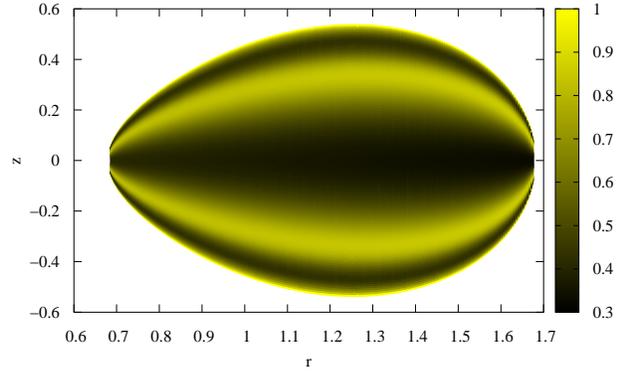}
\caption{
Entropy distribution in the $rz$-plane normalised by its maximum value for Solberg--H{\o}iland-unstable barocline. 
Sequences of $\{q_i\}$ and $\{\beta_i  \}$ are determined by (\ref{q_model}), (\ref{beta_unstable_model_1}) and (\ref{beta_unstable_model_2}).
$q_s=1.5$, $q_e=2$, $N=249$, $\gamma=5/3$, $R_d=1$.
The corresponding baroclinicity of the flow is $G=0.12$.
}
\label{entropy_HZ}
\end{figure}

\begin{figure}
\includegraphics[width=\linewidth]{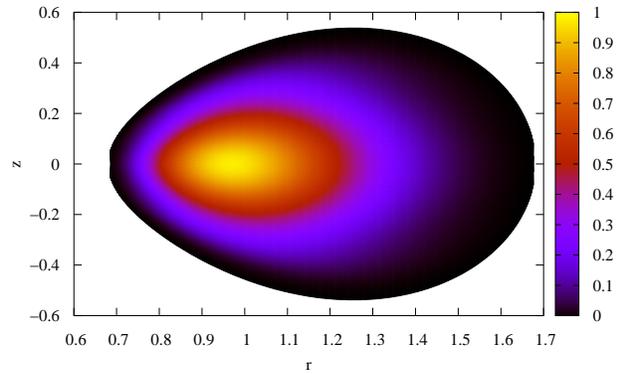}
\caption{
Pressure in the $rz$-plane for Solberg--H{\o}iland unstable barocline normalised by its maximum value for the same parameters as in figure \ref{entropy_HZ}.
}
\label{p_HZ}
\end{figure}

\begin{figure}
\includegraphics[width=\linewidth]{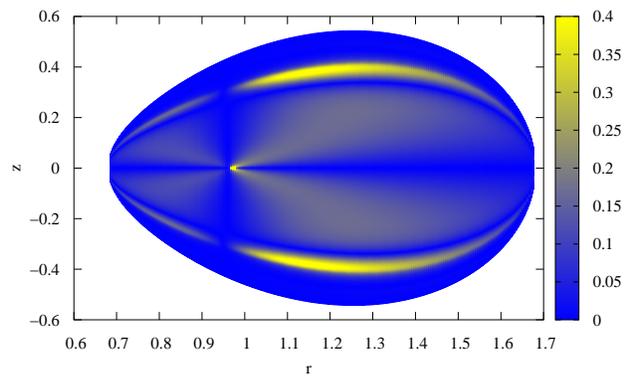}
\caption{
Local baroclinicity in the $rz$-plane for Solberg--H{\o}iland unstable barocline for the same parameters as in figure \ref{entropy_HZ}.
}
\label{local_baroclinicity_HZ}
\end{figure}

As a result, the maximum of entropy appears inside the flow, which leads to destabilization according to the Solberg--H{\o}iland criterion.
The resulting entropy, pressure and local baroclinicity distributions are shown in figures (\ref{entropy_HZ}), (\ref{p_HZ}) and (\ref{local_baroclinicity_HZ}), respectively.
The baroclinicity of the flow is equal to $G=0.12$.

\section{Conclusions}
In this work, we have been able to construct baroclinic tori that are exact solutions of an Eulerian equation in the point-mass Newtonian potential.
A number of modifications to method proposed in \citetalias{amendt-1989} allows us make it free of significant drawbacks and construct the flow which is stable according to the Solberg--H{\o}iland criterion; at the same time it has a significant degree of baroclinicity.

The parameters of the resulting tori can be easily varied by changing the construction parameters.
The baroclinicity of the flow can be increased up to a significant value by increasing the number of the incoming barotropic solutions.
The thickness of the flow is controlled by the parameter $q_e$.

Presented method can be useful in the investigations of the dependence of GSF instability on the background properties such as global baroclinicity.
And since the resulting baroclinic flow have strong radial and vertical entropy gradients, it can be used also in investigations of entropy-driven instabilities as well.

Because of the significant stratification of the resulting baroclinic flow, the regions with radial shear rate $q_r>2$ can be stable.
The stability of such regions with respect to nonaxisymmetric and/or to nonlinear perturbations potentially is an interesting subject to future studies.

\acknowledgements
Author thanks V. V. Zhuravlev for careful reading of manuscript and for meaningful discussion that led to significant changes in the text.
Equipment for the reported study was granted by the M. V. Lomonosov Moscow State University Programme of Development.
Author was supported by grant RSF 14-12-00146.


\begin{thebibliography}{48}
\expandafter\ifx\csname natexlab\endcsname\relax\def\natexlab#1{#1}\fi

\bibitem[{Amendt {et~al.}(1989)Amendt, Lanza, \& Abramowicz}]{amendt-1989}
Amendt, P., Lanza, A., \& Abramowicz, M.~A. 1989, \apj, 343, 437

\bibitem[{{Arlt} \& {Urpin}(2004)}]{arlt-urpin-2004}
{Arlt}, R. \& {Urpin}, V. 2004, \aap, 426, 755

\bibitem[{{Armitage}(2011)}]{armitage-2011}
{Armitage}, P.~J. 2011, \araa, 49, 195

\bibitem[{Balbus \& Hawley(1991)}]{balbus-hawley-1991}
Balbus, S.~A. \& Hawley, J.~F. 1991, \apj, 376, 214

\bibitem[{{Barker} \& {Latter}(2015)}]{barker-latter-2015}
{Barker}, A.~J. \& {Latter}, H.~N. 2015, \mnras, 450, 21

\bibitem[{{Chandrasekhar}(1960)}]{chandrasekhar-1960}
{Chandrasekhar}, S. 1960, Proceedings of the National Academy of Science, 46,
  253

\bibitem[{Frank \& Robertson(1988)}]{frank-robertson-1988}
Frank, J. \& Robertson, J.~A. 1988, \mnras, 232, 1

\bibitem[{Fricke(1968)}]{fricke-1968}
Fricke, K. 1968, Zeitschrift für Astrophysik, 68, 317

\bibitem[{Goldreich {et~al.}(1986)Goldreich, Goodman, \&
  Narayan}]{goldreich-1986}
Goldreich, P., Goodman, J., \& Narayan, R. 1986, \mnras, 221, 339

\bibitem[{Goldreich \& Schubert(1967)}]{goldreich-schubert-1967}
Goldreich, P. \& Schubert, G. 1967, \apj, 150, 571

\bibitem[{{Hur{\'e}} \& {Hersant}(2017)}]{hure-hersant-2017}
{Hur{\'e}}, J.-M. \& {Hersant}, F. 2017, \mnras, 464, 4761

\bibitem[{Johnson \& Gammie(2005)}]{johnson-gammie-2005a}
Johnson, B.~M. \& Gammie, C.~F. 2005, \apj, 626, 978

\bibitem[{Johnson \& Gammie(2006)}]{johnson-gammie-2006}
Johnson, B.~M. \& Gammie, C.~F. 2006, \apj, 636, 63

\bibitem[{{King} {et~al.}(2007){King}, {Pringle}, \& {Livio}}]{king-2007}
{King}, A.~R., {Pringle}, J.~E., \& {Livio}, M. 2007, \mnras, 376, 1740

\bibitem[{{Klahr}(2004)}]{klahr-2004}
{Klahr}, H. 2004, \apj, 606, 1070

\bibitem[{Klahr \& Hubbard(2014)}]{klahr-hubbard-2014}
Klahr, H. \& Hubbard, A. 2014, \apj, 788, 8

\bibitem[{Klahr \& Bodenheimer(2003)}]{klahr-bodenheimer-2003}
Klahr, H.~H. \& Bodenheimer, P. 2003, \apj, 582, 869

\bibitem[{Kojima(1989)}]{kojima-1989}
Kojima, Y. 1989, \mnras, 236, 589

\bibitem[{{Kojima} {et~al.}(1989){Kojima}, {Miyama}, \&
  {Kubotani}}]{kojima-1989b}
{Kojima}, Y., {Miyama}, S.~M., \& {Kubotani}, H. 1989, \mnras, 238, 753

\bibitem[{{Kotko} \& {Lasota}(2012)}]{kotko-lasota-2012}
{Kotko}, I. \& {Lasota}, J.-P. 2012, \aap, 545, A115

\bibitem[{Lesur \& Papaloizou(2010)}]{lesur-papaloizou-2010}
Lesur, G. \& Papaloizou, J. C.~B. 2010, \aap, 513, 12

\bibitem[{{Lipunova}(2015)}]{lipunova-2015}
{Lipunova}, G.~V. 2015, \apj, 804, 87

\bibitem[{{Lipunova} \& {Malanchev}(2016)}]{lipunova-malanchev-2016}
{Lipunova}, G.~V. \& {Malanchev}, K.~L. 2016, ArXiv e-prints

\bibitem[{Lyra(2014)}]{lyra-2014}
Lyra, W. 2014, \apj, 789, 7

\bibitem[{{Marcus} {et~al.}(2016){Marcus}, {Pei}, {Jiang}, \&
  {Barranco}}]{marcus-2016}
{Marcus}, P.~S., {Pei}, S., {Jiang}, C.-H., \& {Barranco}, J.~A. 2016, ArXiv
  e-prints

\bibitem[{{Marcus} {et~al.}(2015){Marcus}, {Pei}, {Jiang}, {Barranco},
  {Hassanzadeh}, \& {Lecoanet}}]{marcus-2015}
{Marcus}, P.~S., {Pei}, S., {Jiang}, C.-H., {et~al.} 2015, \apj, 808, 87

\bibitem[{{Marcus} {et~al.}(2013){Marcus}, {Pei}, {Jiang}, \&
  {Hassanzadeh}}]{marcus-2013}
{Marcus}, P.~S., {Pei}, S., {Jiang}, C.-H., \& {Hassanzadeh}, P. 2013, Physical
  Review Letters, 111, 084501

\bibitem[{{McNally} \& {Pessah}(2015)}]{mcnally-pessah-2015}
{McNally}, C.~P. \& {Pessah}, M.~E. 2015, \apj, 811, 121

\bibitem[{Mukhopadhyay {et~al.}(2005)Mukhopadhyay, Afshordi, \&
  Narayan}]{mukhopadhyay-2005}
Mukhopadhyay, B., Afshordi, N., \& Narayan, R. 2005, \apj, 629, 383

\bibitem[{Nelson {et~al.}(2013)Nelson, Gressel, \& Umurhan}]{nelson-2013}
Nelson, R.~P., Gressel, O., \& Umurhan, O.~M. 2013, \mnras, 435, 2610

\bibitem[{Papaloizou \& Pringle(1984)}]{papaloizou-pringle-1984}
Papaloizou, J. C.~B. \& Pringle, J.~E. 1984, \mnras, 208, 721

\bibitem[{Petersen {et~al.}(2007{\natexlab{a}})Petersen, Julien, \&
  Stewart}]{petersen-2007a}
Petersen, M.~R., Julien, K., \& Stewart, G.~R. 2007{\natexlab{a}}, \apj, 658,
  1236

\bibitem[{Petersen {et~al.}(2007{\natexlab{b}})Petersen, Stewart, \&
  Julien}]{petersen-2007b}
Petersen, M.~R., Stewart, G.~R., \& Julien, K. 2007{\natexlab{b}}, \apj, 658,
  1252

\bibitem[{Razdoburdin \& Zhuravlev(2012)}]{razdoburdin-zhuravlev-2012}
Razdoburdin, D.~N. \& Zhuravlev, V.~V. 2012, Astronomy Letters, 38, 117

\bibitem[{Razdoburdin \& Zhuravlev(2015)}]{razdoburdin-zhuravlev-2015}
Razdoburdin, D.~N. \& Zhuravlev, V.~V. 2015, Physics-Uspekhi, 58, 1031

\bibitem[{{Razdoburdin} \& {Zhuravlev}(2017)}]{razdoburdin-zhuravlev-2017}
{Razdoburdin}, D.~N. \& {Zhuravlev}, V.~V. 2017, \mnras, 467, 849

\bibitem[{{Richard} {et~al.}(2016){Richard}, {Nelson}, \&
  {Umurhan}}]{richard-2016}
{Richard}, S., {Nelson}, R.~P., \& {Umurhan}, O.~M. 2016, \mnras, 456, 3571

\bibitem[{{R{\"u}diger} {et~al.}(2002){R{\"u}diger}, {Arlt}, \&
  {Shalybkov}}]{rudiger-2002}
{R{\"u}diger}, G., {Arlt}, R., \& {Shalybkov}, D. 2002, \aap, 391, 781

\bibitem[{Shakura \& Sunyaev(1973)}]{shakura-sunyaev-1973}
Shakura, N.~I. \& Sunyaev, R.~A. 1973, \aap, 24, 337

\bibitem[{Shen {et~al.}(2006)Shen, Stone, \& Gardiner}]{shen-2006}
Shen, Y., Stone, J.~M., \& Gardiner, T.~A. 2006, \apj, 653, 513

\bibitem[{{Simon} {et~al.}(2012){Simon}, {Beckwith}, \&
  {Armitage}}]{simon-2012}
{Simon}, J.~B., {Beckwith}, K., \& {Armitage}, P.~J. 2012, \mnras, 422, 2685

\bibitem[{{Suleimanov} {et~al.}(2008){Suleimanov}, {Lipunova}, \&
  {Shakura}}]{suleimanov-2008}
{Suleimanov}, V.~F., {Lipunova}, G.~V., \& {Shakura}, N.~I. 2008, \aap, 491,
  267

\bibitem[{Tassoul(1978)}]{tassoul-1978}
Tassoul, J.-L. 1978, Theory of Rotating Stars (Princeton: Princeton Univ.
  Press)

\bibitem[{{Turner} {et~al.}(2014){Turner}, {Fromang}, {Gammie}, {Klahr},
  {Lesur}, {Wardle}, \& {Bai}}]{turner-2014}
{Turner}, N.~J., {Fromang}, S., {Gammie}, C., {et~al.} 2014, Protostars and
  Planets VI, 411

\bibitem[{{Urpin}(2003)}]{urpin-2003}
{Urpin}, V. 2003, \apj, 404, 397

\bibitem[{{Urpin} \& {Brandenburg}(1998)}]{urpin-brandenburg-1998}
{Urpin}, V. \& {Brandenburg}, A. 1998, \mnras, 294, 399

\bibitem[{Velikhov(1959)}]{velikhov-1959}
Velikhov, E.~P. 1959, Sov. Phys. JETP, 9, 995

\bibitem[{Zhuravlev \& Shakura(2007)}]{zhuravlev-shakura-2007b}
Zhuravlev, V.~V. \& Shakura, N.~I. 2007, Astronomy Letters, 33, 673

\end{thebibliography}
\end{document}